\documentclass{WileyMSP-template}
\usepackage[dvipsnames]{xcolor}
\usepackage{ulem}
\usepackage{upgreek}
\usepackage{hyperref}
\begin{document}

\pagestyle{fancy}
\rhead{\includegraphics[width=2.5cm]{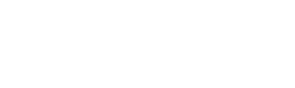}}

\color{red} \textbf{This is the accepted version of the article published open access in Advanced Materials Interfaces: \newline \href{}{https://doi.org/10.1002/admi.202400317} \newline} \color{black}

\title{Stable CoO\textsubscript{2} Nanoscrolls With Outstanding Electrical Properties}

\maketitle
\author{Simon Hettler*}
\author{Kankona Singha Roy}
\author{Raul Arenal*}
\author{Leela S. Panchakarla}

\dedication{This paper is dedicated to the memory of Professor Leela S. Panchakarla, dear supervisor (K.S.R) and colleague, outstanding scientist, who died prematurely.}

\begin{affiliations}
Dr. S. Hettler \\
Instituto de Nanociencias y Materiales de Aragon, CSIC-Universidad de Zaragoza
Laboratorio de Microscopias Avanzadas, Universidad de Zaragoza
Mariano Esquillor Gomez, Zaragoza, 50018, Spain \\
Email Address: hettler@unizar.es

K. S. Roy, Prof. Leela Panchakarla, \\
Department of Chemistry, Indian Institute of Technology Bombay, Mumbai, 400076, Powat, India 

Dr. S. Hettler and K.S. Roy have contributed equally to this work.

Dr. R. Arenal \\
Instituto de Nanociencias y Materiales de Aragon, CSIC-Universidad de Zaragoza
Laboratorio de Microscopias Avanzadas, Universidad de Zaragoza
Mariano Esquillor Gomez, Zaragoza, 50018, Spain
Araid Foundation, Zaragoza, 50018 Spain \\
Email Address: arenal@unizar.es

\end{affiliations}


\begin{abstract}

Layered CoO\textsubscript{2} is of great interest for its promising properties but is meta-stable in its bulk form. CoO\textsubscript{2} was synthesized by converting the quasi-one-dimensional crystal structure of bulk Ca\textsubscript{3}Co\textsubscript{2}O\textsubscript{6} via a hydrothermal treatment. The resulting nanostructures were predominantly nanoscrolls with very thin walls, which exhibit long-term stability. A detailed structural investigation reveals that the CoO\textsubscript{2} is found to crystallize in monoclinic form, similar to the related CaCoO\textsubscript{2}-CoO\textsubscript{2} misfit structure. Individual nanoscrolls are characterized electrically and show a p-type semiconducting nature with a high current-carrying capacity of 4$\cdot$10\textsuperscript{5} A/cm\textsuperscript{2} and an extremely high breakdown voltage of up to 270 kV/cm. The results demonstrate the possibility to stabilize meta-stable materials in low-dimensional forms and a promising application of the nanoscrolls as interconnect in high-voltage electronic circuitry.

\end{abstract}

\section{Introduction}\label{sec1}

\begin{figure}[ht]
    \centering
    \includegraphics[width=0.8\linewidth]{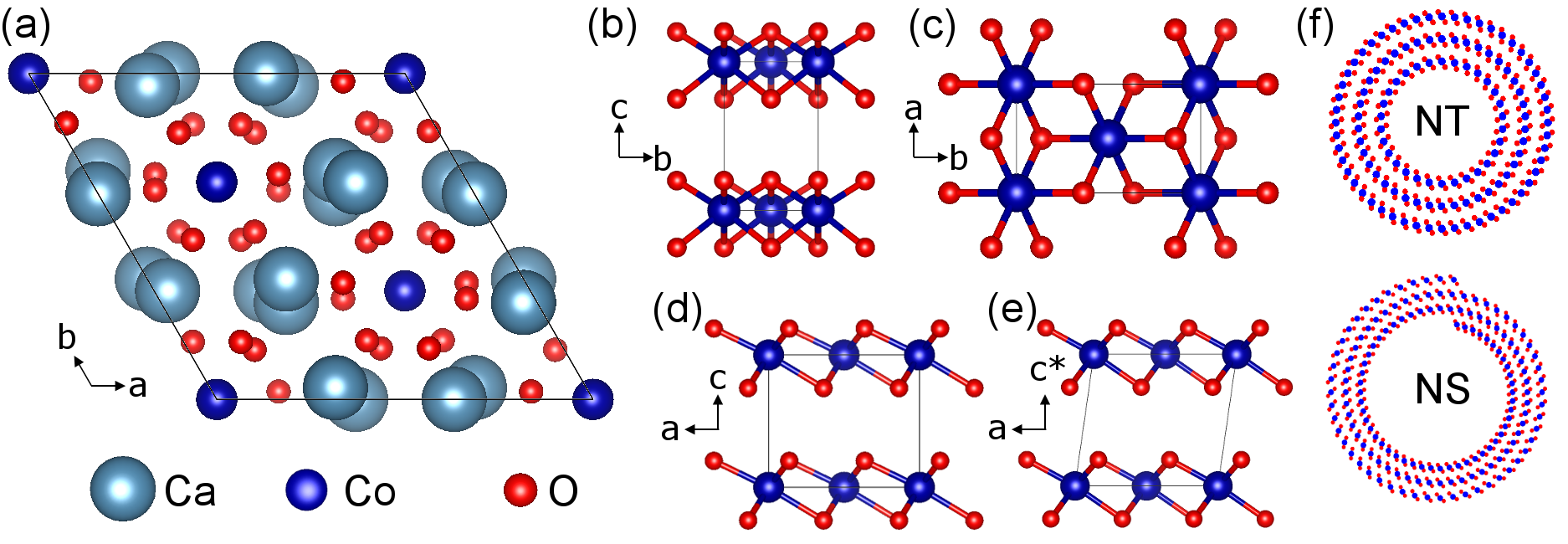}
    \caption{(a) View along c axis of Ca\textsubscript{3}Co\textsubscript{2}O\textsubscript{6} crystal structure reveals the separation of Ca and Co columns. (b-d) View of pseudo-orthogonal CoO\textsubscript{2} crystal structure along a, c and b axis of the unit cell, respectively. (e) View along b axis of monoclinic crystal structure with $\upbeta$~=~98\degree. (f) Schematic representation of a CoO\textsubscript{2} NT and a CoO\textsubscript{2} NS.}
    \label{F1:SketchStru}
\end{figure}

Until the discovery of fullerenes,\cite{Kroto.1985} it was believed that inherent crystal asymmetry was the only driving force to roll layered structures.\cite{Pauling.1930,Serra.2019} Later on it was revealed that flat graphene nanoribbons are unstable and, under the right conditions, spontaneously transform into carbon nanotubes to minimize the energy by elimination of the dangling bonds at the nanoribbons' edges.\cite{Iijima.1991} Tenne et al. created the layered chalcogenides WS\textsubscript{2} as the first inorganic counterparts of carbon nanotubes (CNTs).\cite{Tenne.1992} In contrast to nanotubes, where every wall is formed by an individual, rolled-up monolayer, nanoscrolls (NSs) are made up of a continuous sheet spirally rolled up, resulting in a structure with two open ends (see \textbf{Figure~\ref{F1:SketchStru}f}).\cite{Viculis.2003,Xie.2009,Lai.2016} This gives NSs fundamentally different properties compared to two-dimensional sheets and one-dimensional nanotubes. For instance, carbon NSs can have better mechanical characteristics while retaining the exceptional mobilities from graphene nanosheets.\cite{Wan.2014} Furthermore, NSs may possess a variable diameter along its axis and be readily intercalated, as well as possess a large solvent-accessible surface area. The formation of NSs is driven by the van der Waals interactions of overlapping areas of the layers, which exceed the elastic energy necessary for bending the nanosheet. In the last years, many layered inorganic materials with symmetric and asymmetric structure have been synthesized into nanotubular/nanoscrolled structures, among them MoS\textsubscript{2}, WS\textsubscript{2}, MoSe\textsubscript{2}, and WSe\textsubscript{2} as well as misfit layered compounds.\cite{Serra.2019,Sreedhara.2021} Most of these synthesized structures have a stable corresponding bulk layered structure. 

Several techniques for producing NSs have been developed, including in-situ self-assembly, 1D template-assisted scrolling, inorganic nanoparticle-induced scrolling, and solvent- or polymer-assisted scrolling. \cite{Lai.2016} In these techniques, the starting material would be a 2D-variant or the bulk of the desired layered structure that is subsequently converted into a NS. However, the above-mentioned techniques do not work for the synthesis of NTs/NSs, which have a meta-stable bulk layered structure. Here, we report the synthesis of CoO\textsubscript{2} NSs through hydrothermal conversion of the quasi one-dimensional crystal structure of Ca\textsubscript{3}Co\textsubscript{2}O\textsubscript{6} (Figure~\ref{F1:SketchStru}a),\cite{Fjellvag.1996} similar to the process described in our recent paper on Sr\textsubscript{x}CoO\textsubscript{2}-CoO\textsubscript{2} nanotubes. \cite{Roy.2022} The NSs of CoO\textsubscript{2} are noteworthy, since bulk CoO\textsubscript{2} is meta-stable and is usually obtained only in an intercalated form either with alkali metals or two-dimensional intercalates in misfit-layered compounds. \cite{Mizushima.1980,Tarascon.1999,Masset.2000,Panchakarla.2016} Studies have shown that these CoO\textsubscript{2}-based materials are highly promising due to multiple interesting properties, including unusual superconductivity or ferroelectric instabilities.\cite{Takada.2003,Vaulx.2007,Liang.2021} Meta-stable bulk CoO\textsubscript{2} has been obtained by deintercalation of alkali metals by electrochemical means,\cite{Amatucci.1996,Motohashi.2007} but low-dimensional analogues have not been explored experimentally. In a detailed structural and spectroscopic analysis, the structure of the thin-walled CoO\textsubscript{2} NSs is unraveled. Electrical properties of the NSs have been studied at individual NS level. 
\medskip

\section{Results and Discussion}
\subsection{Structure and composition of CoO\textsubscript{2} nanoscrolls}

Bulk Ca\textsubscript{3}Co\textsubscript{2}O\textsubscript{6} was used as the starting material for the synthesis of the CoO\textsubscript{2} NSs. Figure S1 in supplementary information (SI) shows an X-ray diffraction (XRD) pattern and typical low-resolution scanning electron microscopy (SEM) images of as-synthesized Ca\textsubscript{3}Co\textsubscript{2}O\textsubscript{6}. The XRD pattern coincides well with Ca\textsubscript{3}Co\textsubscript{2}O\textsubscript{6} (ICSD collection code 246281).\cite{Hervoches.2009}

As bulk CoO\textsubscript{2} is meta-stable, experimental data on its crystal structure is scarce.\cite{Tarascon.1999} Figure~\ref{F1:SketchStru}b-d shows the pseudo-orthogonal representation of the hexagonal CoO\textsubscript{2} structure with a=0.483~nm, b=0.282~nm and c=0.424~nm along the three axes. In misfit-layered compounds, CoO\textsubscript{2} is intercalated between sheets of another layered material and the structure adopts a monoclinic structure with $\upbeta$~$\neq$~90\degree,\cite{Leligny.2000,Miyazaki.2002,Pelloquin.2004} being 98\degree  for the (Ca\textsubscript{2}CoO\textsubscript{3})\textsubscript{0.62} - CoO\textsubscript{2} compound.\cite{Miyazaki.2002} Figure~\ref{F1:SketchStru}e shows the view along the b axis of CoO\textsubscript{2} with $\upbeta$~=~98\degree (Crystallography Open Database \#3000496).

\begin{figure}[ht]
    \centering
    \includegraphics[width=1\linewidth]{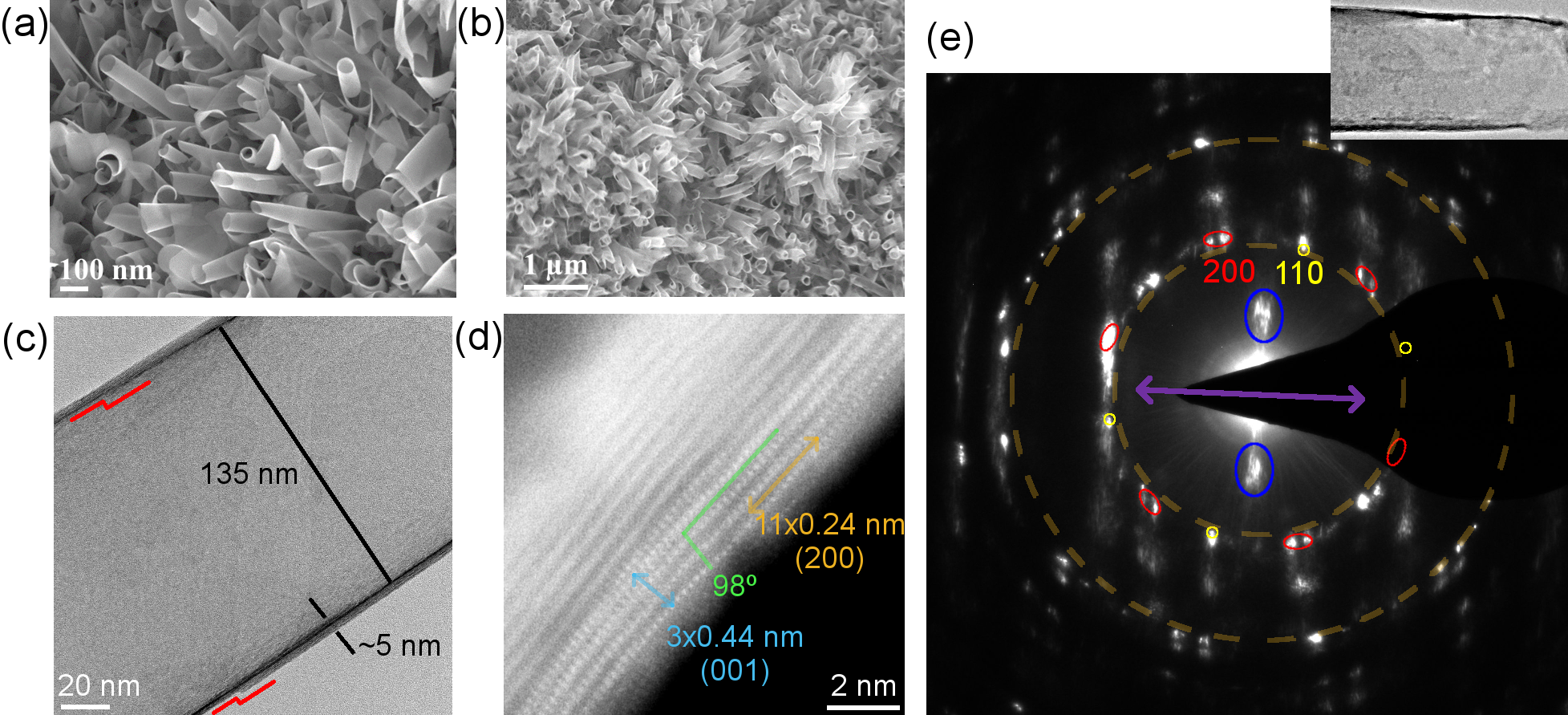}
    \caption{(a,b) SEM, (c) TEM and (d) STEM images of CoO\textsubscript{2} NSs reveal their hollow structure and their extremely thin walls. The stepped red line in (c) confirms the scroll structure.  (e) SAED pattern with different reflections marked, see text for discussion. Pattern is obtained from NS shown in the inset image. Width of pattern is 17.9 nm\textsuperscript{-1} and of inset image 280~nm.}
    \label{F2:EMimages}
\end{figure}

CoO\textsubscript{2} NSs are generated via hydrothermal treatment of Ca\textsubscript{3}Co\textsubscript{2}O\textsubscript{6} in a basic solution (2.5M NaOH) at 220 $\degree$C. The high yield of CoO\textsubscript{2} nanostructures achieved by this approach is visible in the SEM images of the reaction product revealing a predominant presence of thin rolled up sheets, which in many cases form a closed hollow structure (\textbf{Figure~\ref{F2:EMimages}a,b}). The open structures show diameters ranging from 120 to 200 nm and lengths ranging from 500 nm to 1 $\upmu$m. 

Transmission electron microscopy (TEM) confirms the hollow morphology of CoO\textsubscript{2}-NSs and shows that they have extremely thin walls when compared to its diameter. The exemplary NS displayed in Figure~\ref{F2:EMimages}c exhibits a diameter of 135 nm and a wall thickness of 5~nm. A step in the thin wall is indicated by a red line. The fact that this additional sheet is located on opposite sides of the wall, i.e. on the inside of the upper and on the outside of the lower wall, reveals the scrolled structure. Additional images confirming the scrolled structure are shown in Figure S2. High-resolution (HR)TEM images show lattice fringes with a periodicity of 0.44~nm across the wall (Figure S3). This interplanar spacing of 0.44 nm is found in all of the examined NSs and agrees with the c-axis of the crystal structure of CoO\textsubscript{2}. Figure~\ref{F2:EMimages}d shows a high-angle annular dark-field (HAADF) scanning TEM (STEM) image of the wall of a CoO\textsubscript{2}-NS, which as well reveals lattice fringes corresponding to a 0.44~nm layer spacing perpendicular to the wall (light blue double arrow). In the outer layers of the wall, atomic columns are visible along the NS wall with an inter-columnar distance of 0.24~nm, which can be linked to the (200) direction of CoO\textsubscript{2} (orange double arrow in Figure~\ref{F2:EMimages}d). Moreover, the atomic columns are found to encompass an angle of 98\degree (green lines), similar to the monoclinic structure of related misfit-layered compounds but without intercalated second layer.\cite{Pelloquin.2004,Panchakarla.2016,Serra.2019} Both inner and outer layer show a thickness of about 4-5 atomic layers, suggesting that the NS could be made up of a single sheet of CoO\textsubscript{2} with this thickness. The difference in contrast could stem from scrolling, which induces an orientation change of the sheet from inner to outer layer. Figure~S4 shows additional HRSTEM images of  CoO\textsubscript{2} NSs with similar appearance.

To further elucidate the structure of the CoO\textsubscript{2} NSs, we performed selected-area electron diffracion (SAED) on individual NSs. Figure~\ref{F2:EMimages}e displays a SAED pattern of the selected NS, shown in the TEM image (inset of this Figure~\ref{F2:EMimages}e). In the pattern, the c-axis periodicity of about 0.44 nm is confirmed by the corresponding (001) reflections (blue ellipses) lying perpendicular to the tube axis (purple double arrow). Several independent 001 reflections can be observed, which are also considerably smeared. This is consistent with the observed thin walls and indicates a scrolled structure and local variations in the interplanar distance. Higher-order 00l reflections are weak or completely absent, which agrees with expected intensities of such higher orders in the monoclinic structure. Several rings of diffraction spots may be observed, which can be mainly linked to the (200) (0.24 nm) and (020) (0.14 nm) lattice distances of the CoO\textsubscript{2} structure (brown dashed circles in Figure~\ref{F2:EMimages}e). A detailed investigation of the SAED pattern in Figure~\ref{F2:EMimages}e shows that the spots on the (200) ring are actually composed of several individual reflections with slightly different interplanar distances. These additional reflections are consistent with the monoclinic  CoO\textsubscript{2} crystal structure (Figure~\ref{F1:SketchStru}e). In contrast to the orthogonal structure (Co atoms from adjacent layers in c direction sitting directly on top of each other, see Figure~\ref{F1:SketchStru}d), which leads to single (200) and (020) reflections, the monoclinic structure possesses a tilted c-axis ($\upbeta$=98°), which leads to the appearance of additional reflections ((110): 0.243 nm, (-201): 0.224 nm) with distances close to the (200) orientation (0.239 nm). In the SAED pattern (Figure~\ref{F2:EMimages}e), the main reflections can be identified as (200) planes and one set of six reflection spots equally distributed on the circle by 60\degree~has been marked by red ellipses. Each spot is split in two reflections, which encompass an angle of 6\degree, the chiral angle of the NS is thus 3\degree. A second set of (200) reflections is present, which is rotated by 30\degree with respect to the first set (not marked). The (110) reflections have a multiplicity of four and are not equally distributed along the circle but encompass angles of 60\degree~and 120\degree. In Figure~\ref{F2:EMimages}e we marked four (110) reflections (yellow circles), which are rotated by 30\degree with respect to adjacent (200) reflections. The orientations and distances are marked in a schematic representation in Figure S5. Thus, the detailed analysis of the SAED patterns confirms the monoclinic CoO\textsubscript{2} crystal structure observed in HRSTEM imaging. Additional diffraction patterns are shown in Figure~S6. Possible spots coming from a CaCoO\textsubscript{2} layer of the misfit system at 0.36 nm (110) and 0.18 nm (220) are weak or completely absent confirming the CoO\textsubscript{2} structure. 

An X-ray diffraction (XRD) pattern obtained from the synthesis product is shown in Figure~S7, SI. It reveals that the bulk precursor has completely reacted, as no match to the Ca\textsubscript{3}Co\textsubscript{2}O\textsubscript{6} structure is found. Instead, an fcc CoO phase is detected and small CoO crystallites have indeed been observed in TEM analysis (not shown). Additionally, a very broad and weak peak located at the expected position of the 001 reflection of the CoO\textsubscript{2} crystal structure is observed. The absence of a clear signature of the CoO\textsubscript{2} NSs in the XRD pattern is however not surprising and expected, as the crystallite size of the NSs is very small in c direction (sheets with a thickness of a few atoms) and is small and bend in a/b direction due to scrolling, which causes a strong broadening and reduction of the XRD intensity. \cite{Holder.2019}

\begin{figure}[ht]
    \centering
    \includegraphics[width=1\linewidth]{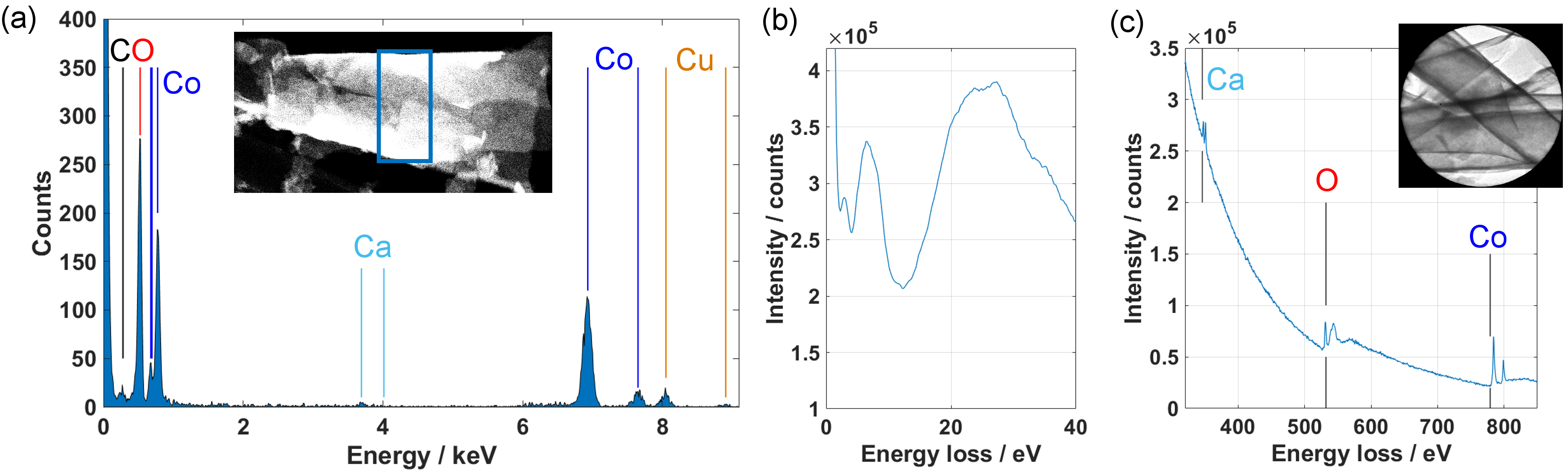}
    \caption{(a) EDS spectrum of an individual NS shows main contributions from Co and O and minor amounts of Ca. Cu and C originate from specimen support structures. Inset TEM image (500~nm width) shows the area from which the spectrum has been obtained. (b,c) EEL spectra of NSs for (b) low-loss region revealing rich plasmonic features and (c) core-loss region revealing presence of Co, O and Ca obtained from the area shown in inset TEM image (500~nm width).}
    \label{F3:DiffSpec}
\end{figure}

Energy-dispersive X-Ray Spectroscopy (EDS) performed in STEM mode on individual NSs reveals the presence of Co, O and a small amount of Ca (see exemplary spectra in \textbf{Figure~\ref{F3:DiffSpec}a} and Figure S8 in SI). The composition is homogeneous throughout the investigated NSs (Figure S8). Together with imaging, the spectroscopic results indicate that Ca is merely a dopant in a structure based on Co and O. Although a quantitative analysis of EDS with theoretical k factors has to be taken with care,\cite{Watanabe.2006} such an analysis of the data acquired on the CoO\textsubscript{2} NS is instructive. The average composition obtained from the acquired spectra is 21-23 at\% Co, 75-78 at\% O and only 0.5-1.2 at\% of Ca, yielding a ratio between O and Co of 3.4, much higher than expected for a CoO\textsubscript{2} structure. The excess O may be linked to surface hydoxyl groups and indeed, when applying a sample cleaning step on the sample dispersed on the grid \cite{Li.2021}, EDS measurements result in a composition much closer to the expected composition (Figure~S9 in SI). In sum, EDS analysis confirms and agrees with a Ca-doped CoO\textsubscript{2} structure.

Electron energy-loss spectroscopy (EELS) in TEM mode has been performed to obtain information on elemental composition and chemistry of the NSs. Figure~\ref{F3:DiffSpec}b,c show exemplary EEL spectra of the low-loss region (Figure~\ref{F3:DiffSpec}b) and the core-loss region including Ca-L, O-K and Co-L edges (Figure~\ref{F3:DiffSpec}c). The low-loss area exhibits two peaks located at 3 eV and 7 eV indicating a rich plasmonic structure of the NSs. The second peak at 7 eV seems to be composed of two individual contributions. The bulk plasmon positioned around 23 eV overlaps with the Ca-M edge (24 eV). The core-loss region may be used to quantify the composition of the investigated region by using calculated scattering cross sections,\cite{Egerton.2011} and the composition is determined to 64 at\% O, 35 at\% Co and 1 at\% Ca, being in good agreement with the EDS results and a predominant CoO\textsubscript{2} structure. The analysis of the energy loss near-edge structure (ELNES) of the Co-L and O-K edges reveals a high energy for the Co-L\textsubscript{3} peak position (781 eV) and a strong pre-peak at 530 eV for the O-K edge, which varies only slightly with the investigated NS (Figure S10). In previous studies, a high energy of the Co-L\textsubscript{3} peak was linked to a higher Co valence,\cite{Gao.2018} while a pre-peak of the O-K edge is highly untypical for conventional cobalt oxides \cite{Barreca.2010,Gao.2018} and resembles more the spectrum of magnetite Fe\textsubscript{3}O\textsubscript{4}. The spectra are similar to the ones observed in a related structure based on strontium cobalt oxide.\cite{Roy.2022} In sum, the EELS analysis suggests a high Co valence, which could be Co\textsuperscript{4+} as expected for CoO\textsubscript{2}.

\subsection{Growth mechanism}

Ca\textsubscript{3}Co\textsubscript{2}O\textsubscript{6} is made up of one-dimensional columns of face-sharing octahedra and trigonal prismatic CoO\textsubscript{6} polyhedra intercalated by Ca atoms that move along the crystallographic c-direction (Figure~\ref{F1:SketchStru}a).\cite{Fjellvag.1996} The distance between two neighbouring cobalt cations in the face-sharing CoO\textsubscript{6} polyhedra is comparable to or slightly less than the distance between two neutral Co atoms in the metal. As a result of the higher electrostatic repulsion between the two closely spaced high valent Co ions, the face sharing arrangement requires a large amount of energy. Ca atoms of the bulk dissolve rapidly in basic solutions during hydrothermal treatment, rendering the face-sharing 1D CoO\textsubscript{6} polyhedra chains unstable. As a result of the missing stabilization by Ca atoms in the lattice, the face-sharing CoO\textsubscript{6} polyhedra change to edge-sharing CoO\textsubscript{6} sheets, which subsequently crystallize into a tubular, scrolled shape. A possible step in the crystal-conversion process could be the Ca\textsubscript{2}CoO\textsubscript{3}-CoO\textsubscript{2} misfit phase, which, however, in the ongoing hydrothermal treatment converts in almost pure CoO\textsubscript{2} NSs. Figure S11 in SI depicts the analysis of the synthesis products after 6h and shows that already at this stage the NSs possess a small wall thickness when coming out of the bulk at the growth points. It is noted that these CoO\textsubscript{2} NSs were found in minute quantity also in the synthesis product of our recent work on strontium-deficient Sr\textsubscript{x}CoO\textsubscript{2}-CoO\textsubscript{2} nanotubes \cite{Roy.2022}. The difference between the Ca and Sr-based approach may thus only be the higher solubility of Ca compared to Sr in the basic solution.

Together with the Ca doping, the curvature in the tubular or scrolled morphologies is a reason for the apparent stabilization of the meta-stable bulk phase of CoO\textsubscript{2}. The thin wall and the limited diameter indicate that this stabilization only occurs for a narrow range of bending radii. The observation of the monoclinic unit cell by STEM imaging and SAED diffraction suggests that this crystal structure is slightly more stable than the orthogonal one. Further studies, including theoretical and computational investigations, will be necessary to explain the stability of the  CoO\textsubscript{2} NSs.

\begin{figure}[ht]
    \centering
    \includegraphics[width=0.6\linewidth]{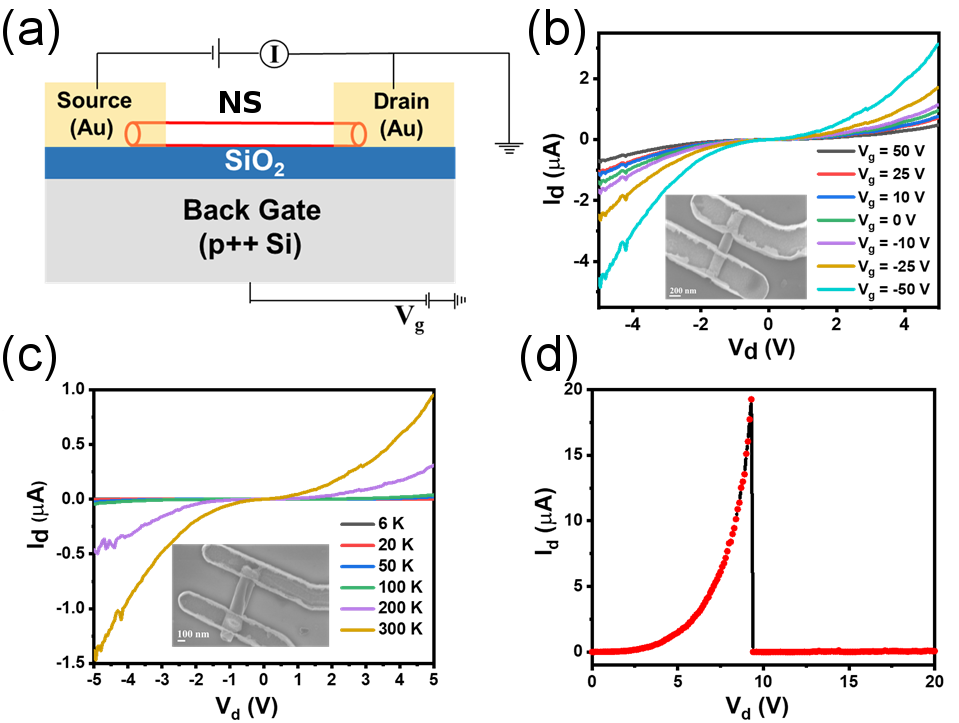}
    \caption{(a) Schematic representation of a two-probe device. (b) Output characteristic of a NS (inset image) for different gate voltages. (c) Temperature dependence of the output characteristics. (d) I-V characteristics until breakdown of the NS depicted in the SEM image in (b). }
    \label{F4:EMeasure}
\end{figure}

\subsection{Electrical properties of CoO\textsubscript{2} nanoscrolls}

In addition to the structural analyses, a dedicated study of the electrical properties of the NSs was performed. \textbf{Figure~\ref{F4:EMeasure}a} denotes the schematic representation of a typical two-probe device, which was realized by electron-beam lithography (see experimental section). Figure~\ref{F4:EMeasure}b represents the output characteristics of an exemplary two-probe device measured under ambient conditions at room temperature. A SEM image of the device is shown as an inset in Figure~\ref{F4:EMeasure}b revealing a channel length of 340 nm and a scroll diameter of 192 nm. The drain voltage (V\textsubscript{d}) is swept from -5 V to +5 V while stepwise increasing the gate voltage (V\textsubscript{g}) from -50 V to +50 V (Figure~\ref{F4:EMeasure}b). With increasing gate voltage (negative), the channel current is increasing, indicating a p-type semiconducting nature of the NSs. Moreover, the current is observed to increase stronger than linearly with the applied drain voltage. Although a contribution from Schottky contacts cannot be completely excluded in the two-probe setup, this non-linear increase very likely occurs due to Joule heating and a resulting increase in electrical conductivity of the semiconducting NSs, especially considering the large maximum current densities between approximately 0.25 and 1.5$\cdot$10\textsuperscript{5}~A/cm\textsuperscript{2} (gray and cyan curves at V\textsubscript{d}~=~-5~V in Figure~\ref{F4:EMeasure}b, respectively). Temperature-dependent I-V characteristics (Figure~\ref{F4:EMeasure}c and Figure S12 in SI) show a decrease in current with decreasing temperature, confirming that the nanotubes are of semiconducting nature. We have measured I-V curves for around 20 devices in a two-probe configuration with channel lengths varying from 300 - 600 nm.

The current-carrying capacity of a device depends on three parameters: the geometry, the electrical resistance and the thermal conductivity, which determines the induced temperature rise. Even when the device architecture is uncomplicated, the interplay between resistance and temperature rise is a complex phenomenon and depends on many factors. Single CoO\textsubscript{2} NSs can withstand electrical currents of 20-25~$\upmu$A under ambient conditions before failure (at V\textsubscript{g} = 0~V) (Figure~\ref{F4:EMeasure}d and Figure S13 in SI), resulting in a maximum current density of up to 4$\cdot$10\textsuperscript{5}~A/cm\textsuperscript{2}, assuming a wall thickness of 5~nm. Although lower than hBN coated carbon nanotubes (CNTs) \cite{Huang.2015}, this value of high current density surpasses metals like Au, Pt and W,\cite{W.H.Preece.1883} and is comparable to compounds like TaSe\textsubscript{3}, Nb-Ti, PbSe or SnO\textsubscript{2}. \cite{Chernyj.1991,Empante.2019,Qin.2020} The measured maximum current density is most probably only a lower approximation as the current does not saturate with applied bias.

Interestingly, the breakdown voltage per channel length is surprisingly high for these devices (9.2 V/340~nm = 270 kV/cm for Figure~\ref{F4:EMeasure}d), as high as for hBN-capped CNTs.\cite{Huang.2015} The breakdown would occur when the electric field is sufficiently strong enough to pull electrons from the material and ionize it. Due to the high oxidation state of cobalt, the ionization process is difficult, resulting in high stability under high voltages. The breakdown is thus more likely to be a result of Joule heating and/or electromigration. The SEM image in Figure S13d (in SI) shows the NS after device failure indicating that the device breakdown occurred not inside the metal contacts but inside the NS, close to the contact. This observation, together with the rounded shape after breakdown (Figure S13d) suggests that defect-induced Joule heating caused the melting and breakdown of the NS. When looking at the breakdown power per channel length (P/L), the CoO\textsubscript{2}-NSs can withstand $\cong$7.5 W/cm (Figure S13), which outperforms most used interconnect candidates, i.e. Cu (polycrystalline)\cite{Rathmell.2011} and stands close to BN-capped quasi-one-dimensional TaSe\textsubscript{3} ($\cong$6.8 W/cm).\cite{Stolyarov.2016} A measurement under vacuum or capping the device with a dielectric layer (hBN, Al\textsubscript{2}O\textsubscript{3}, SiO\textsubscript{2} etc.) could suppress the electromigration process, further increase the breakdown power and the current-carrying capacity.\cite{Huang.2015} 

It is worth to mention that no leakage through the substrate was observed as proven by a measurement without NS (Figure S14a in SI). Furthermore, the devices are stable for over 16 months at ambient conditions (Figure S14b in SI), suggesting a high device and NS stability, making a use of the CoO\textsubscript{2} NSs, for example as interconnects, in real-life applications promising.

\section{Conclusions}

We synthesized bulk meta-stable CoO\textsubscript{2} in a stable form of nanoscrolls (NSs) by a crystal conversion method. The crystal structure of the NSs is investigated by electron microscopy imaging and diffraction and CoO\textsubscript{2} is found in a monoclinic structure with $\upbeta$~=~98\degree. Spectroscopic techniques are consistent with a Co oxidation state of 4+. The stabilization of the CoO\textsubscript{2} NSs, which are long-term stable ($>$ 1 year) at ambient conditions, is understood to be created mainly by the curvature of the sheets and this stabilization seems to occur only for a narrow range of scroll diameters and wall thicknesses (bending radii). The employed crystal conversion method is therefore promising to realize more meta-stable compounds yet to be synthesized. Electrical properties of individual CoO\textsubscript{2} NSs were determined and they exhibit a high current-carrying capacity (4×10\textsuperscript{5} A/cm2) and an extraordinary high electrical breakdown voltage (270~kV/cm) and power per unit channel length ($\cong$7.5 W/cm). These outstanding performances and the stability of these CoO\textsubscript{2} NSs makes their use as potential building blocks for high-power electronic applications highly promising.

\section{Experimental}

\threesubsection{Synthesis of CoO\textsubscript{2} nanoscrolls}
The NSs were synthesized by dispersing 20 mg of Ca\textsubscript{3}Co\textsubscript{2}O\textsubscript{6} bulk material in 2.5M NaOH solution via sonication for 15 min. The dispersion is transferred to an air-tight autoclave, which was placed in a constant-temperature oven. The oven was heated to 220 $\degree$C with a 5 $\degree$C/min rate and kept at that temperature for 12h before letting cool down to room temperature overnight. Milli-Q water was used to wash the product until it became pH neutral. The final product was achieved by drying at 60 $\degree$C overnight.

\threesubsection{Analysis techniques}
The synthesis product was characterized using various techniques. Scanning electron microscopy (SEM) was performed with a ZEISS GeminiSEM 560. Most of the transmission electron microscopy (TEM) studies, including selected-area electron diffraction (SAED) and electron energy-loss spectroscopy (EELS) in TEM mode (Gatan Image Filter (GIF) Tridiem, acceptance angle of 11.9 mrad) were performed using an image-corrected Titan Cube microscope (Thermo Fisher Scientific) working at 300 kV. High-resolution scanning TEM (HRSTEM) imaging analyses were conducted in a probe-corrected Titan Low-Base microscope (Thermo Fisher Scientific) operated at 300 kV, using the high-angle annular dark-field (HAADF) detector (convergence/acceptance angle 25/48 mrad). Energy-dispersive X-ray spectroscopy (EDS) using an Oxford Instruments Ultim Max TLE 100 detector was performed in the same instrument. Some complementary TEM measurements were performed in a Themis 300 G3 (Thermo Fisher Scientific). The Ca\textsubscript{3}Co\textsubscript{2}O\textsubscript{6} and CoO\textsubscript{2} powders were analyzed by X-ray diffraction (XRD) using a PANalytical diffractometer Empyrean system with CuK\textsubscript{$\alpha$} radiation.

\threesubsection{Device fabrication and electrical measurements}
A NS powder dispersion was prepared by sonication for 1 min of the powder in iso-propyl alcohol (IPA) (HPLC grade) and then spin-coated (3000 pm, 1 min) on a pre-patterned Si (high p-doping) /SiO2 wafer. Individual NSs were marked and mapped in SEM. For the following electron-beam lithography process of source and drain electrodes (Raith 150 two), the substrate with NSs was coated with e-beam resist EL9 and PMMA 950K 2\% was spin-coated on the substrate with NSs before e-beam expose. Development of the resist was achieved by dipping in a 1:3 MIBK/IPA mixture for 30s followed by immediate immersion in IPA for 10 s. Sputtering of metal contacts ( 5 nm Cr/ 50 nm Pt/ 50 nm Au) was performed with an Orion sputter machine followed by an Ar bias cleaning step to make sure that all PMMA is removed. Pt (6.35 eV) acts as the work function material and enables ohmic contact. Cr facilitates good adhesion and a swift lift-off process. For this lift-off step, the substrate was kept in acetone (5h) followed by washing with acetone, IPA and drying under N\textsubscript{2} flow. Due to the small length of the NSs, we fabricated two-probe devices, which were characterized with a semiconductor characterization system (Proxima Keysight B1500A) at room temperature under ambient conditions. Measurements were also performed at low temperatures (down to 6 K) under vacuum using a Lakeshore CRX-4K cryostat. Ampacity was determined at ambient conditions. NSs-free areas were used as test structure to guarantee that the SEM imaging and subsequent metal deposition do not lead to a contacting of source and drain.

\medskip
\textbf{Supporting Information} \par 
Supporting Information is available from the Wiley Online Library or from the author.

\medskip
\textbf{Acknowledgements} \par 
Dear late Prof. L.S.P. would like to acknowledge Science and Engineering Research Board under the Department of Science and Technology (DST-SERB), Government of India for funding (EMR/2016/003594). K.S.R. acknowledges the Department of Chemistry, Department of Physics and Department of Materials Engineering and Materials Science, Indian Institute of Technology (IIT) Bombay and Industrial Research and Consultancy Center (IRCC), Sophisticated Analytical Instrument Facility (SAIF), Centre of Excellence in Nanoelectronics (CEN) IIT Bombay and Fund for Improvement of S\&T Infrastructure (FIST) for all the facilities provided. K.S.R. sincerely thanks Gayatri Vaidya at IIT Bombay for helpful assistance and discussion about the fabrication process. S.H. and R.A. acknowledge funding from the European Union’s Horizon 2020 research and innovation programme under the Marie Sklodowska-Curie grant agreement No. 889546 and from the Spanish MICIU (project grant PID2019-104739GB-100/AEI/ 10.13039/501100011033) . R.A. acknowledges funding from the Government of Aragon (project DGA E13-23R) and from the European Union H2020 programs ‘‘ESTEEM3’’ (823717) and ‘‘Graphene Flagship’’ (881603) as well as from the Spanish MICIU by the “Severo Ochoa” Programme for Centres of Excellence in R\&D (CEX2023-001286-S MICIU/AEI/10.13039/501100011033). Most of the TEM measurements were performed in the Laboratorio de Microscopias Avanzadas (LMA) at the Universidad de Zaragoza (Spain). The authors thank G. Antorrena for help with XRD acquisition.

\medskip

\bibliographystyle{MSP}
\bibliography{biblio}

\begin{figure}[h]
\textbf{Table of Contents}\\
\medskip
  \includegraphics{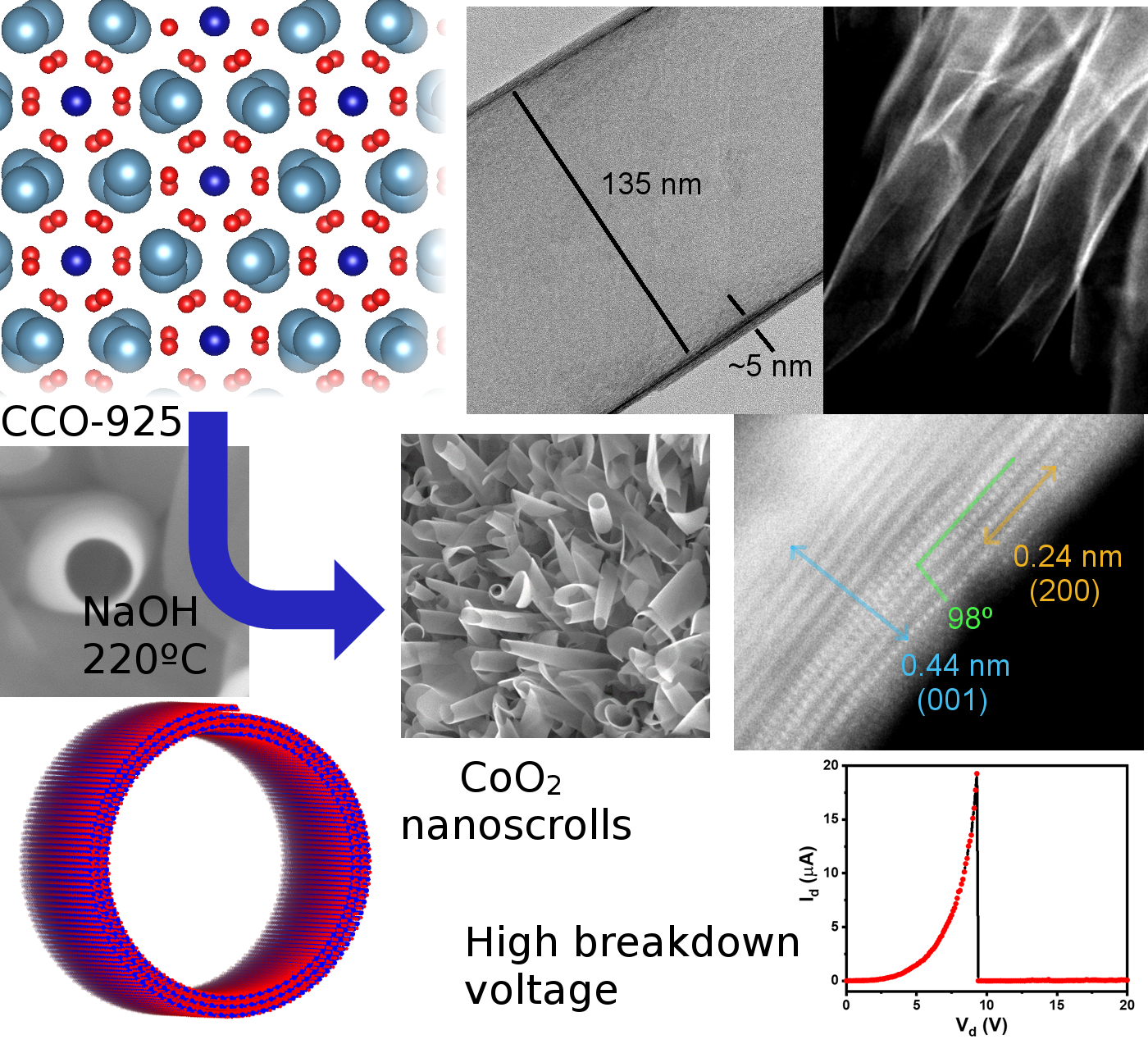}
  \medskip
  \caption*{Layered cobalt dioxide, unstable in bulk form, was synthesized in the form of long-term stable nanoscrolls with narrow walls by crystal conversion from a bulk precursor with a quasi 1D crystal structure. Structure and electrical properties of the nanoscrolls were studied in detail. The nanoscrolls show an extraordinarily high breakdown voltage.}
\end{figure}

\end{document}